\documentclass[10pt,letterpaper]{article}
\usepackage{opex3}
\usepackage{subeqn}
\usepackage{url}
\usepackage[normalem]{ulem}
\newcommand{\be}{\begin{equation}}
\newcommand{\ee}{\end{equation}}
\newcommand{\bea}{\begin{eqnarray}}
\newcommand{\eea}{\end{eqnarray}}

\newcommand{\sech}{\mbox{sech}}
\begin{document}


\title{Coupled Manakov equations in multimode fibers with strongly coupled groups of modes}


\author{Antonio Mecozzi,$^{1,*}$ Cristian Antonelli,$^1$ and Mark Shtaif$^2$}
\address{
$^1$ Department of Physical and Chemical Sciences, University of L'Aquila, 67100 L'Aquila, Italy  \\
$^2$ School of Electrical Engineering, Tel Aviv University, Tel Aviv, Israel 69978}
\email{$^*$antonio.mecozzi@univaq.it} 

\begin{abstract}
We derive the fundamental equations describing nonlinear propagation in multi-mode fibers in the presence of random mode coupling within quasi-degenerate groups of modes. Our result generalizes the Manakov equation describing mode coupling between polarizations in single-mode fibers. Nonlinear compensation of the modal dispersion is predicted and tested via computer simulations.
\end{abstract}


\section{Introduction} \noindent Optical spatial division multiplexing in multi-mode optical fibers (MMFs) has been recently attracting enormous attention as it is considered to be a promising solution for the current capacity crunch. Optical nonlinearity is one of the most fundamental aspects of fiber propagation and its effect on the ability to transmit information is of utmost importance. {In a recent paper we have considered the case of nonlinear propagation in multi-mode fibers where all excited modes experience strong random mode coupling \cite{Mecozzi}. In realistic fiber-communications scenarios, it often happens that while some modes have similar wavenumbers and therefore experience strong random coupling, their coupling to other modes  (whose wavenumbers are significantly different) is much weaker. The case of the weakly guiding step-index optical fiber \cite{Gloge} is a typical example for this situation as the linearly polarized LP$_{mn}$ modes contain 2-fold and 4-fold degeneracies. In this paper we show that in this framework nonlinear multi-mode propagation is accurately described in terms of \textit{coupled} generalized Manakov equations (ME) \cite{Manakov} {\color{black} and give the expression for the coupling coefficients}. The advantage of this description {\color{black} over that based on standard coupled nonlinear Schr\"{o}dinger equations (NLSE)} is that it significantly reduces the number of relevant nonlinear parameters and hence simplifies the entire description of nonlinear propagation. It also forms an excellent starting point for analytical studies of nonlinear effects in fiber-optic transmission \cite{OurArchive}. In order to validate the accuracy of the coupled ME, we show that the solitary solutions that are predicted on their basis can indeed be observed in direct simulations of the complete model which is based on the coupled NLSE.}


\section{Analysis and results}
We start from the most general form of the coupled NLSE for MMFs  \cite{Poletti} and analytically introduce the effect of random mode coupling within groups of degenerate modes. The assumption under which our theory works is that the correlation length characterizing the effect of random mode coupling is shorter \cite{Mecozzi} than the length-scale of the nonlinear interaction between pulses. This regime is strictly satisfied in the case of the single-mode fiber supporting two polarization modes \cite{Marcuse} and since the mechanism of mode coupling is caused by perturbations which are of the same nature,  extension to larger mode numbers (in the absence of explicit experimental data on the nature of perturbations) is plausible. This assumption is also consistent with reported experimental results \cite{Ryf,Salsi}.
Our formalism allows a seamless generalization of all results known in the context of nonlinear propagation in single-mode fibers to the multi-mode case.

We consider a fiber with $N$ spatial modes, such that the total number of scalar modes including polarizations is $2N$. Defining a $2N$-dimensional electric field vector $\vec E$ such that its components are the complex amplitudes of the various modes, the coupled NLSE describing field propagation through the MMF is given by \cite{Poletti}
\bea \frac{\partial \vec E}{\partial z} &=& i \mathbf B^{(0)} \vec E  - \mathbf B^{(1)} \frac{\partial \vec E}{\partial t} - i  \frac{\mathbf B^{(2)}} 2 \frac{\partial^2 \vec E}{\partial t^2}  + i \gamma \sum_{jhkm} C_{jhkm} E_h^* E_k E_m \hat e_j, \label{10} \eea
where the dimensionless constants $C_{jhkm}$ depend on the details of the spatial mode profiles. We find it convenient to introduce $\gamma = \omega_0 n_2 / c A_\mathrm{eff}$, which is  identical to the usual nonlinearity coefficient appearing in the scalar NLSE of a single mode fiber, where $n_2$ is the Kerr coefficient of glass, $c$ is the speed of light in vacuum, and $A_\mathrm{eff}$ is the effective area of the fundamental mode at central frequency $\omega_0$. Here, $\mathbf B^{(i)}(z)$, $i=0,1,2$ are $2N\times2N$ Hermitian matrices of components \cite{Antonelli} $\beta_{jm}^{(i)}(z)$ and we use $\hat e_j$ with $j=1,\dots, 2N$ to denote the set of complex orthogonal unit vectors used to represent the electric field, namely $\vec E = \sum_j E_j \hat e_j$. We now assume that the modes are divided between two groups denoted by $a$ and $b$. The two groups are characterized by distinctly different propagation constants, but the modes within each of the groups are nearly degenerate. We will denote the number of degenerate modes in groups $a$ and $b$ by $N_a$ and $N_b$, respectively, such that $N_a+N_b=N$. We assume the existence of only two such groups only for the simplicity of presentation, since the final results can be readily expanded such that any number $M$ of groups of degenerate modes can be accounted for. We denote by $\beta_a$ and by $\beta_b$ the wavenumbers of the two groups of modes, and assume that the difference between them is large enough to ensure that mode coupling between the groups is negligible on the scale of the nonlinear length. In the case of the LP$_{01}$ and LP$_{11}$ mode groups of a step index fiber, this property has been verified both computationally \cite{Antonelli} and experimentally \cite{Ryf,Salsi}.
The coupled generalized multi-component \cite{Makhankov} ME, which are the main result of this Letter, are given by \cite{OurECOC2012}:
\bea \frac{\partial \vec E_a}{\partial z} &=& i \beta_a \vec E_a - \beta_a' \frac{\partial \vec E_a}{\partial t} - i  \frac{\beta_a''} 2 \frac{\partial^2 \vec E_a}{\partial t^2}  + i \gamma \left(\kappa_{aa} |\vec E_a|^2 + \kappa_{ab} |\vec E_b|^2 \right)\vec E_a, \label{20} \\
\frac{\partial \vec E_b}{\partial z} &=& i \beta_b \vec E_b - \beta_b' \frac{\partial \vec E_b}{\partial t} - i  \frac{\beta_b''} 2 \frac{\partial^2 \vec E_b}{\partial t^2}  + i \gamma \left(\kappa_{ba} |\vec E_a|^2 + \kappa_{bb} |\vec E_b|^2 \right)\vec E_b, \label{25} \eea
where $\vec E_a$ and $\vec E_b$ are the vectors describing the modes in groups $a$ and $b$, respectively, in a reference frame that accommodates for the unitary evolution induced by the linear coupling within each group, and their respective number of components is $2N_a$ and $2N_b$. The quantities $\beta'_{a,b}$ and $\beta_{a,b}''$ are the group velocity and the group velocity dispersion terms of the two groups of modes, and the parameters $\kappa_{u v}$ are generalized self-phase modulation and cross-phase modulation (XPM) coefficients given by
\be \kappa_{u v} = \!\! \sum_{k,m}\sum_{j \in u}\sum_{h \in v} C_{jhkm}\frac{\delta_{hk}\delta_{jm}+\delta_{hm}\delta_{jk}}{(2 N_u)(2 N_v + \delta_{v u})}, \label{30} \ee
where $\delta_{xy}$ is the Kronecker delta function. In (\ref{30}), each of the indices $u$ and $v$ takes the values $a$ and $b$ depending on which of the coefficients in Eqs. (\ref{20}) and (\ref{25}) is evaluated. For example, if $u=a$ and $v=b$, then in the summation the index $j$ runs over all values corresponding to the modes in group $a$ and the index $h$ runs over all the indices corresponding to the modes in group $b$. The indices $k$ and $m$ in the summation span all the modes. Because of the symmetry of the coefficients \cite{Poletti} $C_{jhkm} = C_{hjkm}$ we have $\kappa_{u v} = \kappa_{v u} = \mu$.

The coupled ME (\ref{20}) and (\ref{25}) have a significant complexity advantage over the generic NLSE (\ref{10}),  which is of relevance both in the numerical and analytical contexts. In the numerical context, we note that in a fiber supporting $M$ groups of degenerate modes, there are $M$ self-phase modulation coefficients $\kappa_{jj}$ and $M(M-1)/2$ cross-phase modulation coefficients, one for every pair of groups. The overall number of nonlinear coefficients thus scales quadratically with the number of groups. This is to be compared with the number of coefficients $C_{jhkm}$ appearing in Eq. (\ref{10}), which scales as the fourth power of the total mode number $N$.  In the analytical context, we note that while Eq. (\ref{10}) in its generality does not display any simple qualitative feature (in general, solitary pulses are not supported), we will show in what follows that Eqs. (\ref{20}) and (\ref{25}) have the characteristic behavior of two coupled nonlinear equations that support soliton solutions.

The derivation of Eqs. (\ref{20}) and (\ref{25}) starts from the notion that the nonlinear interaction between the two groups of modes can be averaged with respect to the random linear evolution of the fields within each group. That is because in the regime of strong coupling the linear evolution within the groups is significantly faster than the nonlinear evolution. As the statistics of the linear evolution are isotropic, the averaged nonlinear coupling is bound to be isotropic as well, and the nonlinear coupling between the groups can only depend on the fields amplitudes, not on their orientations. In this case, the nonlinear term in the averaged equations representing the evolution of the field in mode $u$ must have the form $i\gamma \left(\kappa_{uv} |\vec E_v|^2 + \kappa_{uu} |\vec E_u|^2 \right)\vec E_u$, where the first and second terms correspond to the effects of cross-phase and self-phase modulation, respectively. Consistency with the last term of Eq. (\ref{10}) implies the validity of the following substitution
\be \sum_{k,m} \sum_{j \in u} \sum_{h \in v} C_{jhkm} E_h^* E_k E_m \hat e_j \longmapsto \kappa_{uv} |\vec E_v|^2 \vec E_u. \label{20a} \ee
Scalar multiplication of the right-hand-side of (\ref{20a}) by $\vec E_u$ is equivalent to scalar multiplication of the left-hand-side by the extension of $\vec E_u$ to the $2N$ dimensional vector representation used in Eq. (\ref{10}). Averaging with respect to the orientations of $\vec E_u$ and $\vec E_v$ yields
\be \kappa_{uv} = \sum_{k,m} \sum_{j \in u} \sum_{h \in v} C_{jhkm} \frac{\mathcal{E} \left[ E_h^* E_k E_m E_j^*\right]}{|\vec E_u|^{2}|\vec E_v|^2},\label{eq101}\ee
where the symbol $\mathcal{E}$ denotes statistical averaging. The derivation of Eq. (\ref{30}) from Eq. (\ref{eq101}) is presented in the appendix. 
{\color{black} Equations (\ref{20})--(\ref{30}) have been derived in \cite{Mumtaz} for the special case $N_a=N_b=1$ using a different approach. This special case represents a situation in which none of the spatial modes couple with each other, so that strong coupling can only be attributed to polarizations.}

\section{Numerical validation}
{In order to validate Eqs.(\ref{20})--(\ref{30}), one would have to demonstrate that for any arbitrary excitation, propagation according to these equations leads to the same field evolution as propagation according to Eq. (\ref{10}).  In principle this can be done with the kind of excitation commonly encountered in telecom applications, but a far more convenient and elegant choice would be to take advantage of the fact that Eqs. (\ref{20})--(\ref{30}) can be shown to have analytical solitary solutions with predictable properties. In the case of solitary pulses, nonlinearity plays a much more pronounced  role than it does in most typical communications scenarios. Therefore, by demonstrating that the same pulses are observed in the numerical solution of Eq. (\ref{10}), the accuracy of Eqs. (\ref{20})--(\ref{30}) can be verified.} In the numerical studies discussed in what follows we assume a weakly guiding step-index fiber supporting the LP$_{01}$ and LP$_{11}$ mode groups. Since mode LP$_{11}$ is two-fold degenerate, this scenario is represented in our notation by setting $N_a=1$ and $N_b=2$, where $a$ and $b$ refer to modes LP$_{01}$ and LP$_{11}$, respectively.
{When only one of the groups is excited (in a particular mode and polarization), it can be shown by direct substitution that a hyperbolic-secant pulse of the form}
\be E_{u} = a_{u} \sech \left( \frac{t-T_{u}}{\tau_{u}} \right)\exp\left(-i\omega_u t\right) \label{80} \ee
{is an exact solitary solution of Eqs. (\ref{20})--(\ref{30}), where $a_u$, $T_u$, $\tau_u$ and $\omega_u$ are parameters representing the amplitude, the central position, the temporal width, and the center frequency---defined as the offset from the carrier frequency---of the optical pulse and where $a_u$ and $\tau_u$ satisfy the relation $\tau_u a_u=\sqrt{|\beta_u''|/(\gamma\kappa_{uu})}$ and $\omega_u = 0$, with $u=a$ or $u =b$. In the case where both mode-groups are excited, it can be shown that solitary solutions {\color{black} still exist and} can be very well approximated by pulses whose functional form is identical to (\ref{80}), with 
the parameters $a_u$, $T_u$, $\tau_u$ and $\omega_u$ (for $u=a,b$) being analytically expressible \cite{OurArchive}}.
\begin{figure}[t!]
\begin{centering}
\includegraphics[width=.9\columnwidth]{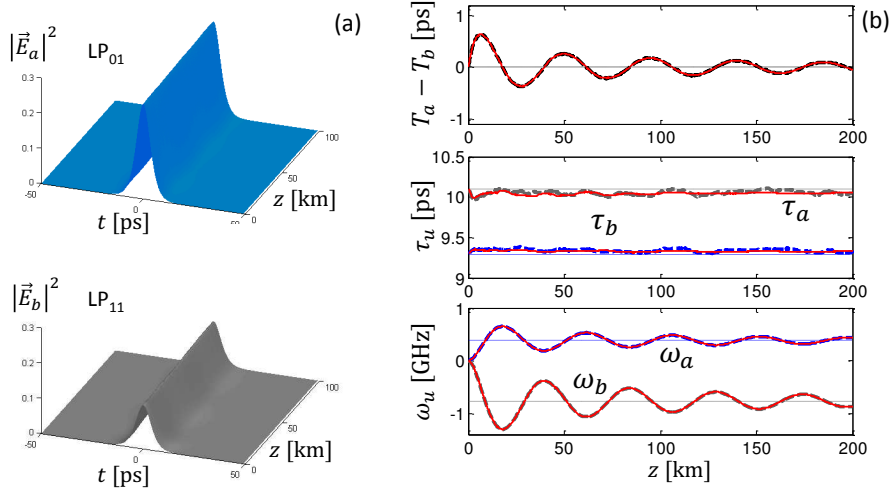}
\caption{(a) Three-dimensional plot of the power envelope of the two modes obtained from Eq. (\ref{10}). The fact that solitary evolution is demonstrated validates the coupled ME, on the basis of which solitary evolution was predicted. (b) Timing difference $T_a - T_b$, center frequencies (denoted by $\omega_a$ and $\omega_b$), and pulse-widths $\tau_a$ and $\tau_b$ as a function of the propagation distance $z$. The black-dashed curves represent the solution of Eq. (\ref{10}) and the solid-red by the solution of the coupled Manakov equations (\ref{20})--(\ref{30}). These two curves are hardly differentiable in the resolution of the figure. The horizontal dashed-dotted lines show the analytical steady state values obtained in \cite{OurArchive}.} \label{fig1}
\end{centering}
\end{figure}

{In our simulations,
we excited one polarization mode in each of the LP$_{01}$ and LP$_{11}$ mode groups with a hyperbolic secant pulse as in (\ref{80}). The parameters $a_u$, $T_u$ and $\tau_u$ were set to their predicted steady state values, but the center frequencies at the launch position were both set to zero  (while their steady state values would be different from each other \cite{OurArchive}). The pulses were propagated based on the full NLSE model (\ref{10}) with the expectation that they should converge to the solitary solution.}
We assumed a core radius of $7.5\mu$m, a core refractive index of $1.46$, a refractive index step of $9.7 \times 10^{-3}$ {(similar to \cite{SillardECOC11})}, dispersion coefficients $\beta_{a}'' = - 25$ps$^2$/km and $\beta_{b}'' = - 29$ps$^2$/km, and a differential group delay between the two groups of $\beta_{a}' - \beta_{b}' = 0.2 $ps/km. The nonlinear coefficient in our computations was $n_2 = 2.6 \times 10^{-20}$m$^2$W$^{-1}$ and the effective area for the fundamental mode was $A_\mathrm{eff} = 126\mu$m$^2$, corresponding to $\gamma \simeq 0.835 $ W$^{-1}$km$^{-1}$, $\kappa_{aa} \simeq 0.89$, $\kappa_{bb} \simeq 0.76$, $\mu \simeq 0.895$. A fundamental difficulty in the simulations is to accommodate the very large wavenumber difference between the two groups, which is estimated to be $\beta_{a} - \beta_{b} \simeq 9.153\times10^{3} $m$^{-1}$. This value corresponds to a beat-length smaller than a millimeter and since the overall length of the simulated fiber should be in the tens or hundreds of kilometers range, simulations become prohibitively inefficient. In order to bypass this difficulty, the wavenumber that we used in the simulations was scaled down by a factor of 10,000, resulting in a beat-length of the order of 10m. The integration-step in the split-step solution of the coupled NLSE was correspondingly limited not to exceed 0.1m.

Figure \ref{fig1}(a) shows the power envelope of the two pulses propagating in the LP$_{01}$ and LP$_{11}$ modes (denotes as $|E_a|^2$ and $|E_b|^2$, respectively). These plots were obtained by solving Eq. (\ref{10}) numerically. The solitary evolution of the two pulses, which is predicted on the basis of the coupled ME (\ref{20})--(\ref{30}) \cite{OurArchive}, validates the accuracy of the latter. In Fig. \ref{fig1}(b) we show several attributes of the pulses propagating in the two groups of modes. The initial  temporal widths of the two pulses used in the simulation were $\tau_a = 9.3$ps and $\tau_b = 10.1$ps, and the pulse energies in the LP$_{01}$ and LP$_{11}$ modes were $5$pJ and $2.5$pJ, respectively.
In each of the plots in Fig. \ref{fig1}(b) the dashed and the solid curves represent the results of the coupled NLSE (\ref{10}) and of the coupled ME (\ref{20})--(\ref{25}) respectively. As is evident from the figure, the difference between those two curves is hardly noticeable indicating the accuracy of the coupled Manakov model. The flat dotted lines represent the {\color{black} parameters of the} steady state solution
{\color{black} described above} to which the numerical solutions are clearly seen to converge.  
{\color{black} In the steady state solution, the initial difference of group velocity of the two groups of modes is compensated by a self-generated frequency shift of the two solitons \cite{OurArchive}. This dynamics is identical to soliton trapping observed in single-mode birefringent fibers \cite{Gordon_self_trapping}}.

\section{Conclusions}
To conclude, we have derived a set of coupled ME describing the nonlinear evolution of the electric field in a MMF in the presence of random mode coupling. These equations highlight that the nonlinear evolution in MMFs in the presence of random mode coupling is characterized by a small number of parameters that scales with the square of the number of non-degenerate groups of modes. This is in striking contrast with the the number of parameters required to describe the nonlinear interaction in the absence of random mode coupling, which scales with the total number of modes to the fourth power. In order to demonstrate the validity of the coupled ME, we showed that an explicit numerical solution of the full coupled NLSE (\ref{10}) produces the solitary solutions predicted by the coupled Manakov model.

\section*{Appendix: Equivalence of Eqs. (\ref{30}) and (\ref{eq101}).} \noindent Our goal is to perform the average appearing in Eq. (\ref{eq101}) with respect to the orientations of the filed vectors $\vec E_u$ and $\vec E_v$, which can be assumed to have uniform distributions. To this end we introduce auxiliary random vectors $\vec X_u$ and $\vec X_v$, with $2N_u$ and $2 N_v$ complex, statistically independent components. The real and imaginary parts of each component $X_i$ of $\vec X_u$ and $\vec X_v$ are statistically independent standard Gaussian variables having zero-mean and unit variance. Using known properties of Gaussian vectors, it can be shown that the term multiplying the coefficients $C_{jhkm}$ in Eq. (\ref{eq101}) is equal to $Q_{hkmj} = \mathcal{E}\left[ X_h^* X_k X_m X_j^*\left||\vec X_u|^2,|\vec X_v|^{2}\right.\right]/(|\vec X_u|^{2}|\vec X_v|^{2})$. Multiplying both sides by $|\vec X_u|^2 |\vec X_v|^2$ and performing another average (with respect to the square modulus $|\vec X_u|^2$ and $|\vec X_v|^2$), we find that
\be \frac{\mathcal{E}\left[ X_h^* X_k X_m X_j^*\right]}{\mathcal{E}\left[|\vec X_u|^{2} |\vec X_v|^{2}\right]}=\frac{\delta_{hk}\delta_{jm}+\delta_{hm}\delta_{jk}}{(2 N_u)(2 N_v + \delta_{v u})}, \ee
where we have made use of standard properties of Gaussian variables.



\begin{thebibliography}{99}

\bibitem{Mecozzi} A. Mecozzi, C. Antonelli, and M. Shtaif,  ``Nonlinear propagation in multi-mode fibers in the strong coupling regime,'' Opt. Express \textbf{20}, 11673--11678 (2012).

\bibitem{Gloge} D. Gloge, ``Weakly guiding fibers,'' Appl. Opt. \textbf{10}, 2252--2258 (1971).

\bibitem{Manakov} S. V. Manakov, ``On the theory of two-dimensional stationary self-focusing of electromagnetic waves,'' Sov. Phys. JETP \textbf{38}, 248--253 (1974).

\bibitem{OurArchive} A. Mecozzi, C. Antonelli, and M. Shtaif,  ``Soliton trapping in multimode fibers with random mode coupling," arXiv:1207.6506v2 [physics.optics] (2012).

\bibitem{Poletti} F. Poletti and P. Horak, ``Description of ultrashort pulse propagation in multimode optical fibers,'' \josab \, \textbf{25}, 1645--1654 (2008).

\bibitem{Marcuse} D. Marcuse, C. R.  Menyuk, and P. K. A. Wai,  ``Application of the Manakov-PMD equation to studies of signal propagation in optical fibers with randomly varying birefringence,'' J. Lightwave Technol. \textbf{15}, 1735--1746 (1997).

\bibitem{Ryf} R. Ryf, S. Randel, A. H. Gnauck, C. Bolle, A. Sierra, S. Mumtaz, M. Esmaeelpour, E. C. Burrows, R. Essiambre, P. J. Winzer, D. W. Peckham, A. H. McCurdy, and R. Lingle, ``Mode-division multiplexing over 96 km of few-mode fiber using coherent 6 $\times$6 MIMO processing," J. Lightwave Technol. \textbf{30}, 521--531 (2012).

\bibitem{Salsi} M. Salsi, C. Koebele, D. Sperti, P. Tran, H. Mardoyan, P. Brindel, S. Bigo, A. Boutin, F. Verluise, P. Sillard, M. Astruc, L. Provost, and G. Charlet, ``Mode division multiplexing of 2 $\times$ 100Gb/s channels using an LCOS based spatial modulator,'' J. Lightwave Technol. \textbf{30}, 618--623 (2012).

\bibitem{Antonelli} C. Antonelli, A. Mecozzi, M. Shtaif, and P. J. Winzer,  ``Stokes-space analysis of modal dispersion in fibers with multiple mode transmission,'' Opt. Express \textbf{20}, 11718--11733 (2012).

\bibitem{Makhankov} V. G. Makhan'kov and O. K. Pashaev, ``Nonlinear Schr\"{o}dinger equation with noncompact isogroup,'' Theor. Math. Phys. \textbf{53}, 55--67 (1982).

\bibitem{OurECOC2012} A. Mecozzi, C. Antonelli, and M. Shtaif,  ``Optical nonlinearity in multi-mode fibers with random mode coupling,'' Proceedings of ECOC 2012, Paper P.1.11 (2012).

\bibitem{Mumtaz} S. Mumtaz, R. J. Essiambre, and G. P. Agrawal, ``Nonlinear propagation in multimode and multicore fibers: generalization of the Manakov equations," arXiv:1207.6645v1 [physics.optics] (2012).

\bibitem{SillardECOC11} P. Sillard, M. Bigot-Astruc, D. Boivin, H. Maerten, and L. Provost ``Few-mode fiber for uncoupled mode-division multiplexing transmissions,'' Proceedings of ECOC 2011, Paper Tu.5.7 (2011).

\bibitem{Gordon_self_trapping} M. N. Islam, C. D. Poole, and J. P. Gordon, ``Soliton trapping in birefringent optical fibers,'' \ol \, \textbf{14}, 1011--1013 (1989).








\end{thebibliography}
\end{document}